\documentclass[12pt, letterpaper]{article}

\RequirePackage{packages} 
\usepackage[bottom]{footmisc}
\usepackage{soul}
\usepackage{subcaption}

\title{How to Disrupt a Market\thanks{%
Authors are listed alphabetically. Correspondence should be addressed to Edoardo Gallo (\texttt{edo@econ.cam.ac.uk}) and Jonathan Lusthaus (\texttt{jonathan.lusthaus@sociology.ox.ac.uk}). 
\par
We are grateful to seminar participants at the University of Cambridge, University of Melbourne, University of East Anglia and University of Oxford, as well as conference participants at the Australia New Zealand Workshop in Experimental Economics and the Workshop on UK Digital Economics for their helpful comments. 
\par
This study was funded by a grant from the Cisco University Research Program Fund, a corporate advised fund of Silicon Valley Community Foundation (ID: 1376716). Federico Varese's contribution to this paper was supported by funding from the European Research Council (ERC) under the European Union's Horizon 2020 research and innovation program (Grant agreement No.~101020598 -- CRIMGOV).}}

\usepackage{authblk}

\author[1,2]{Edoardo Gallo}
\author[1,3]{Rebecca Heath}
\author[4,5]{Jonathan Lusthaus}
\author[4,6]{Federico Varese}

\affil[1]{\emph{Faculty of Economics, University of Cambridge}}
\affil[2]{\emph{Magdalene College, University of Cambridge}}
\affil[3]{\emph{Fitzwilliam College, University of Cambridge}}
\affil[4]{\emph{Department of Sociology, University of Oxford}}
\affil[5]{\emph{Oxford School of Global and Area Studies, University of Oxford}}
\affil[6]{\emph{Centre for European Studies and Comparative Politics, Sciences Po}}
\date{Last updated: \today }

\begin{document}

\maketitle

\begin{abstract}

Market design research in economics naturally focusses on how to improve market efficiency. Our objective here is exactly the opposite - how to design interventions that make a market less efficient. Our research is inspired by the growth of illicit markets online where reducing their efficiency may reduce societal harm. Using a web-based experiment, we find that a partial disruption to delivery is an effective method to decrease market efficiency. The decrease is borne by sellers who sell fewer goods and have lower earnings. A consequence of a disruption to delivery, however, is an increase in market concentration because it facilitates the emergence of a dominant seller. In contrast, we find that attacks on seller ratings are ineffective at reducing market efficiency. This study paves the way for evidence-based, causally driven investigations to aid policies to disrupt cybercrime and other illicit markets.
    
\end{abstract}

\clearpage

\section{Introduction}

Market design has been one of the most successful applications of economic theory in a broad number of areas, from the assignment of spectrum frequencies for mobile operators \parencite{McMillan2003} to matching people in organ donations \parencite{Roth2004}. The central focus of this research agenda is how to achieve market efficiency to maximise collective welfare. However, in recent years, the emergence of new types of markets has highlighted a key challenge to the universality of this goal. Cybercriminals have developed their own web of shadowy virtual marketplaces where they trade in, for example, stolen data, malware kits, money laundering services and beyond \parencite{lusthaus2018,Motoyama2011,Christin2013,Dupont2017}. Aside from this novel development, illicit markets have existed prior to the Internet, including for drugs, prostitution, humans, and stolen goods \parencite{galeotti2025homo,andreas2025illicit}. These markets illustrate that, in certain contexts, maximising collective welfare requires disrupting the market, i.e. decreasing its efficiency.

Despite the abundance of research in the market design literature, the question of `bad markets' and how best to disrupt such markets has been largely ignored. While we adopt a broad approach that likely applies to a range of illicit markets, we take cybercrime as a particular inspiration for our study. Given the scale of cybercrime alone, developing methods to disrupt these kinds of markets presents a pressing challenge for policy. The Federal Bureau of Investigation (FBI) receives approximately 750,000 cybercrime complaints per year \parencite{fbi2023}. These attacks range from mundane intrusions to attacks on the Federal Reserve. The FBI has calculated that the reported crimes alone cost victims more than \$12.5 billion in 2023 \parencite{fbi2023}. Despite policing efforts, traditional enforcement strategies, such as site takedowns or the arrest of a small number of important members, face significant limitations. Sites that are taken down reform under different names, key players remain hidden behind online anonymity, and many top cybercriminals reside in hubs, such as Russia, Nigeria or China, where law enforcement cooperation may be compromised due to geopolitical reasons or corruption \parencite{PublicAffairs2023,Lusthaus2017}.

The core objective of this paper is to explore how to disrupt a market and decrease its efficiency. Specifically, we examine the impact of two types of attack, a rating attack and delivery attack, on market efficiency. There are considerable policy implications for innovative policing strategies targeting both new criminal markets online, as well as conventional criminal markets in the offline world. To our knowledge, this paper is the first empirical study of how to disrupt a market using an online experiment. Our key finding is that a partial disruption to the delivery of goods is an effective way to disrupt a market because it worsens the efficiency of the market and has a negative impact on sellers' earnings.

\section{Markets of Asymmetric Information and Strategies for Disruption}

Motivated by applications to cybercrime, we focus on markets characterised by asymmetric information. Cybercrime marketplaces are akin to illicit eBays and are characterised by quality uncertainty and identity uncertainty \parencite{Yip2013}. Quality is difficult to validate: photos and screenshots can be easily doctored, malware has been found with backdoors targeting its users, and stolen credit card data can be sold to multiple customers despite any promises of exclusivity. Identities are obscured through using nicknames, proxy services, the Onion Router (Tor) network and the use of cryptocurrencies, making it difficult to differentiate between ``rippers'' and ``honest'' cybercriminals \parencite{franklin,herley2009nobody}, a problem that is pervasive in illegal markets \parencite{gambetta1993sicilian}.

Cybercriminals have developed means to, at least partially, mitigate this uncertainty \parencite{Lusthaus2012,DcaryHtu2013}. Reputation may be built up through repeated interactions, abidance to social norms and signalling past criminal actions. Some cybercriminal marketplaces have formal reputation systems that allow customers to rate transactions and/or leave qualitative feedback on sellers \parencite{DcaryHtu2013}. This is consistent with a well-established body of theoretical and empirical research on reputation in markets with asymmetric information. Theoretically, reputational mechanisms can sustain high-quality production when sellers are deterred from deviating through the threat of punishment \parencite{klein,Shapiro1983} or when buyers use past behaviour to infer seller type \parencite{Kreps1982,Milgrom1982}, thereby avoiding the ``lemons market'' effect \parencite{Akerlof1970}. Empirically, studies on e.g., eBay find that positive seller reputations increase both buyers' willingness to pay \parencite{Resnick2006,Houser2006} and sales \parencite{He2022}. As such, this research may carry broader implications for more traditional online markets, such as eBay and Amazon, that have similar structural characteristics.

\section{Approach and Methodology}

The direct or indirect manipulation of reputational information is an obvious starting point to develop strategies to disrupt a market. There has been some limited discussion of disruptive interventions within the computer science and criminology literatures, particularly in relation to the so-called Slander and Sybil attacks, but this has been conceptual and without empirical investigation. In short, the Slander attack involves leaving fake reviews for sellers, while the Sybil attack entails agents posing as vendors and defaulting on sales \parencite{franklin,DcaryHtu2013,Hutchings2016b}. While not fully fleshed out in the literature, there are numerous attacks that are similar to, or are variations of, Slander and Sybil. In this paper, we investigate the effects of two such manipulations on the efficiency of the market. First, we examine a rating attack that affects the numerical score buyers give to a seller following a transaction. There is a related literature on noise in rating systems in cooperation games, which finds mixed effects. Noise encourages forgiveness and reduces retaliation in bilateral transactions \parencite{fudenbergy}, but it reduces co-operation in network settings \parencite{Gallo2022}.

The second market disruption manipulation strategy is a delivery attack that ``intercepts'' or degrades a good before the buyer receives it. This is similar in spirit to the Sybil attack without requiring the creation of seller accounts, or the associated sale of illegal goods, by law enforcement. Examples of the delivery attack include a bank cancelling a database of stolen credit cards following their sale, cyber intelligence agents exploiting bugs in a piece of malware, and law enforcement damaging the infrastructure that underpins its botnet to degrade or destroy the purchased product/service. Failure to deliver a good as promised has a reputational impact on the seller, and it may therefore affect the efficiency of the market by reducing the likelihood of future trades with the affected seller.

There are a number of practical and ethical challenges tied to academics disrupting cybercriminals in ``the wild'', and accurately measuring the impact of those disruptions. In order to investigate the effect of these two attacks in a controlled setting, we ran an online experiment using the oTree platform \parencite{Chen2016}, with participants recruited from Amazon Mechanical Turk. In total, we collected data on 14 complete groups for each treatment condition, comprised of 392 participants.

\section{\label{sec::design} Experimental Design}

In the experiment, participants are randomly assigned to groups of seven, with each group forming a ``marketplace''. There are three sellers and four buyers in each marketplace and participants play for 20 market rounds. Group and role assignment stay the same throughout the experiment.

\begin{figure}[t!]
    \includegraphics[width=\textwidth]{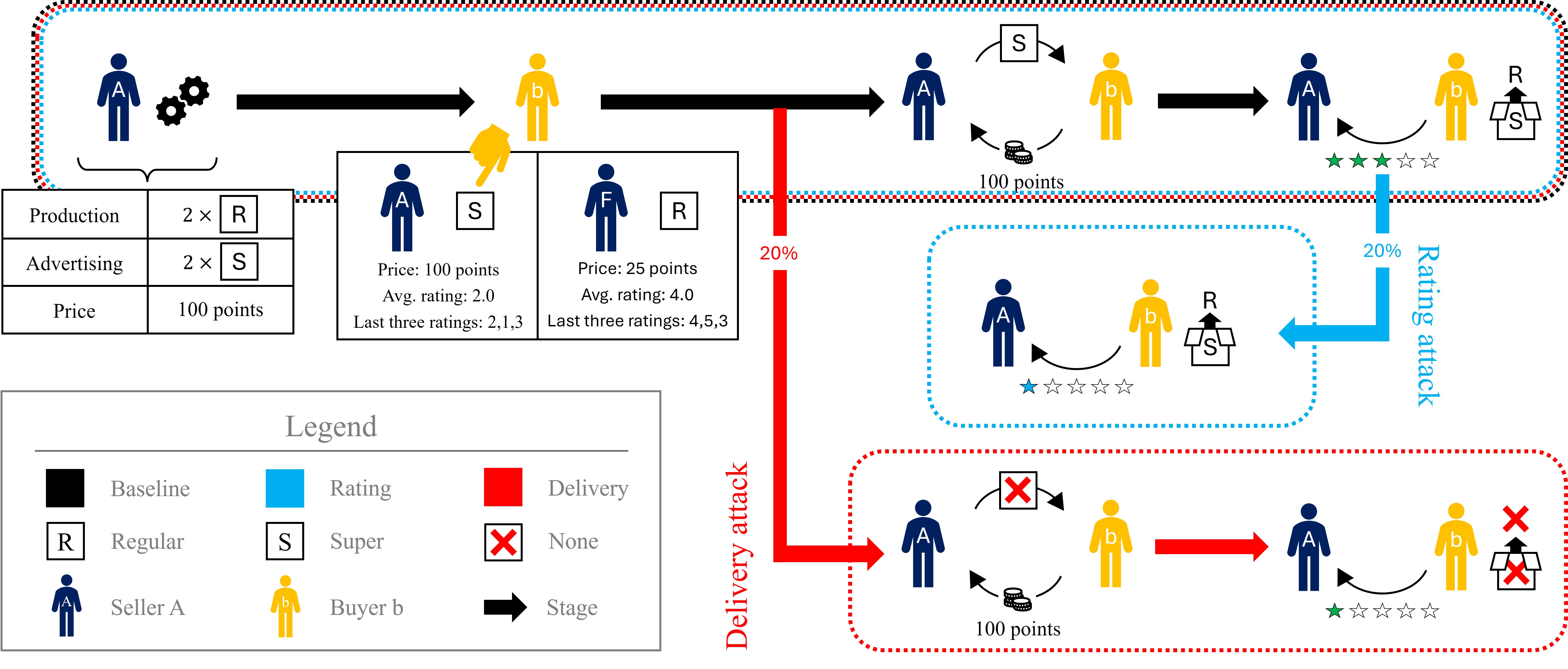}
    \caption{Summary of experimental design. \emph{\footnotesize In the Baseline treatment, there is no disruption: the buyer receives exactly the good sent by the seller, and the seller receives exactly the rating submitted by the buyer. In the Rating treatment, the good is delivered as intended, but the rating system is disrupted - there is a 20\% probability that each rating is replaced with a different, random rating. In the Delivery treatment, the delivery mechanism is disrupted - there is a 20\% probability that the buyer receives nothing, regardless of the seller's production decision. Since the buyer is unaware that the seller has been affected by the delivery attack, the attack may also have a reputational effect.} \label{fig::experimental-design-cyber-one}}
\end{figure}

\Cref{fig::experimental-design-cyber-one} outlines the order of decisions in each market round, including example decisions by Seller A and Buyer b. Dotted lines indicate the treatment conditions.

Starting from the top left of \cref{fig::experimental-design-cyber-one}, each round begins with sellers simultaneously deciding: the quantity and quality of goods to produce; the quantity and quality of goods to advertise; and a price per unit. Each seller can produce 0, 1 or 2 goods, and each good can be of high (super) or regular quality with the additional constraint of one quality level per round (e.g. two super goods, or one regular good, and so on, but never one of each). Production costs are 50 and 10 points for super and regular goods respectively, and the cost of producing the second unit is twice the first unit. In the example, Seller A produces two goods of regular quality but advertises two goods of super quality, illustrating that sellers can advertise more goods and/or goods of a different quality than those produced.

Following this, buyers are randomly allocated into a purchase order. As shown in \cref{fig::experimental-design-cyber-one}, each buyer in turn sees the advertised goods that remain available and decides whether to purchase one of these goods or not make a purchase. Each good is indexed by price, advertised quality, seller's identification code, seller's last three ratings and seller's average rating from all previous transactions. In the \cref{fig::experimental-design-cyber-one} example, two goods are available to Buyer b: a good of super quality advertised by Seller A for 100 points, and a good of regular quality advertised by Seller F for 25 points. Seller F has a higher average rating (4.0) than Seller A (2.0), although both received the same most recent rating (3.0). Buyer b chooses to purchase the good from Seller A for 100 points.

As the top right part of \cref{fig::experimental-design-cyber-one} shows, after a purchase is made, the real quality level is revealed to the buyer who is then asked to rate the seller on a 1 (``Very poor'') to 5 (``Very good'') numerical scale. A buyer obtains 150 and 30 points for any received genuine super and regular goods respectively. In the example, Buyer b receives a good of regular quality, reflecting Seller A's production decision, and submits a rating of 3.0 reflecting that the delivered good was of a lower quality than advertised. A market round ends after all buyers have made their decisions. \\

\noindent \emph{Treatments.} The design described above is our Baseline (B) treatment. We explore three treatment variations, Delivery (D), Rating (R) and Combined (C). We employ a $2\times2$ full-factorial between-subjects design, and therefore each participant only experiences one treatment.

In the delivery treatment, for each purchase decision, there is a 20\% chance that the buyer receives nothing regardless of whether the advertised good was actually produced or not. The dotted red box in \cref{fig::experimental-design-cyber-one} shows that after a delivery attack the Buyer b does not receive the good, and, in the example, submits a rating of 1.0. Notice that Buyer b is not informed that the missing delivery was caused by a delivery attack. In the rating treatment, for each purchase decision there is a 20\% chance that the buyer's rating is replaced with a different, random rating. The dotted blue box in \cref{fig::experimental-design-cyber-one} shows a rating attach where Buyer b's submitted rating of 3.0 is replaced with a rating of 1.0. Finally, the combined treatment includes both interventions with the rating and delivery attacks occurring with independent probabilities.

\section{Results}

We present data for the final 10 market rounds, owing to significant time trends in the data (see Appendix for further details).

\subsection{The Delivery Intervention Reduces Market Efficiency}

\begin{figure}[htb!]
    \includegraphics[width=\textwidth]{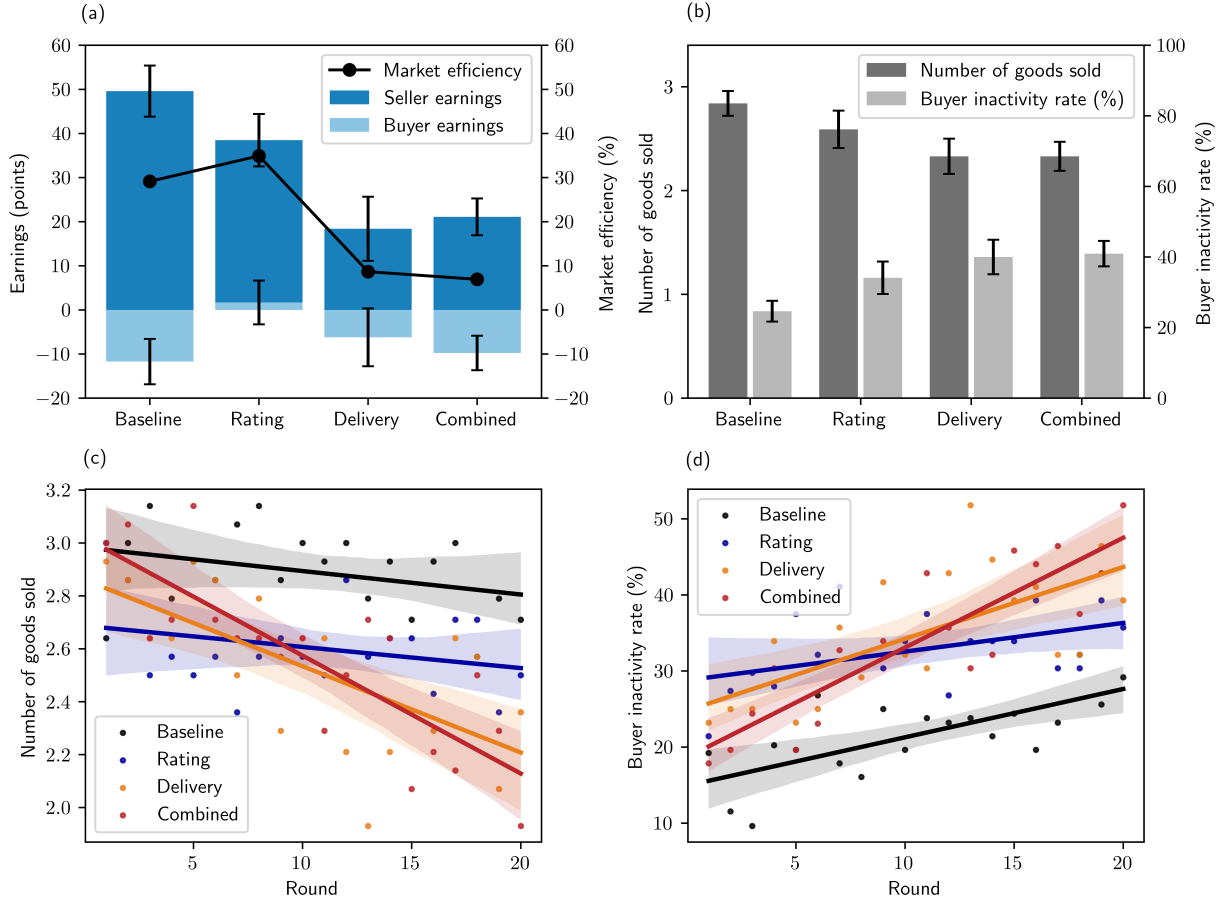}
    \caption{The effect of the interventions on market aggregates. \emph{\footnotesize (a) Market efficiency, sellers' earnings and buyers' earnings averaged over the last 10 market rounds. Bars indicate $\pm$ standard error. (b) Number of goods sold and buyer inactivity rate averaged over the last 10 market rounds. Bars indicate $\pm$ standard error. (c) The number of goods sold over the market rounds, plotted for each intervention. Lines are linear projections based on treatment averages with the 95\% confidence interval shaded. (d) Buyer inactivity rate over the market rounds, plotted for each intervention. Lines are linear projections based on treatment averages with the 95\% confidence interval shaded.}}\label{fig::efficiency_graphs}
\end{figure}

\noindent \cref{fig::efficiency_graphs}{\color{blue}a} shows average market efficiency and earnings across treatments. The delivery intervention lowers market efficiency, defined as the proportion of the maximum gains from trade realised by participants. In particular, market efficiency in the delivery and combined interventions is, respectively, 70\% and 76\% lower than in the baseline (MW, $p<0.01$ for D and C). The decrease in market efficiency is borne by sellers - sellers' earnings in the delivery and combined are, respectively, 63\% and 57\% lower than in the baseline (MW, $p<0.01$ for D and $p<0.001$ for C). Buyers' earnings are, instead, unchanged across treatments. The rating intervention has no impact on market efficiency. 

The decrease in market efficiency remains significant after adjusting the maximum gains from trade for the direct effect of good seizures (refer to Appendix).  The additional decrease in market efficiency is driven by a reduction in the number of goods sold, shown in \cref{fig::efficiency_graphs}{\color{blue}b}. The number of goods sold in the delivery and combined treatments is 18\% lower than in the baseline (MW, $p<0.05$ for D and $p<0.01$ for C). \cref{fig::efficiency_graphs}{\color{blue}c} displays the evolution of the number of goods sold over time divided by treatment. The only two treatments for which there is a significant time trend of a decrease in the number of goods sold over time are the delivery and combined (W, $p<0.05$ for D and $p<0.01$ for C). \cref{fig::efficiency_graphs}{\color{blue}d} illustrates the same trend from the demand side by plotting the evolution of buyer inactivity rate - the proportion of buyers that choose not to purchase a good even though there is at least one available to purchase. The buyer inactivity rate in the delivery and combined is 62\% and 66\% higher than in the baseline (MW, $p<0.05$ for D and $p<0.01$ for C). There is a significant time trend of an increase in inactivity rate in the delivery and combined treatments (W, $p<0.01$ for D, and $p<0.001$ for C).

\subsection{The Delivery Intervention Increases Market Concentration}

\begin{figure}[t!]  
    \centering
    \resizebox{\textwidth}{!}{
        \includegraphics{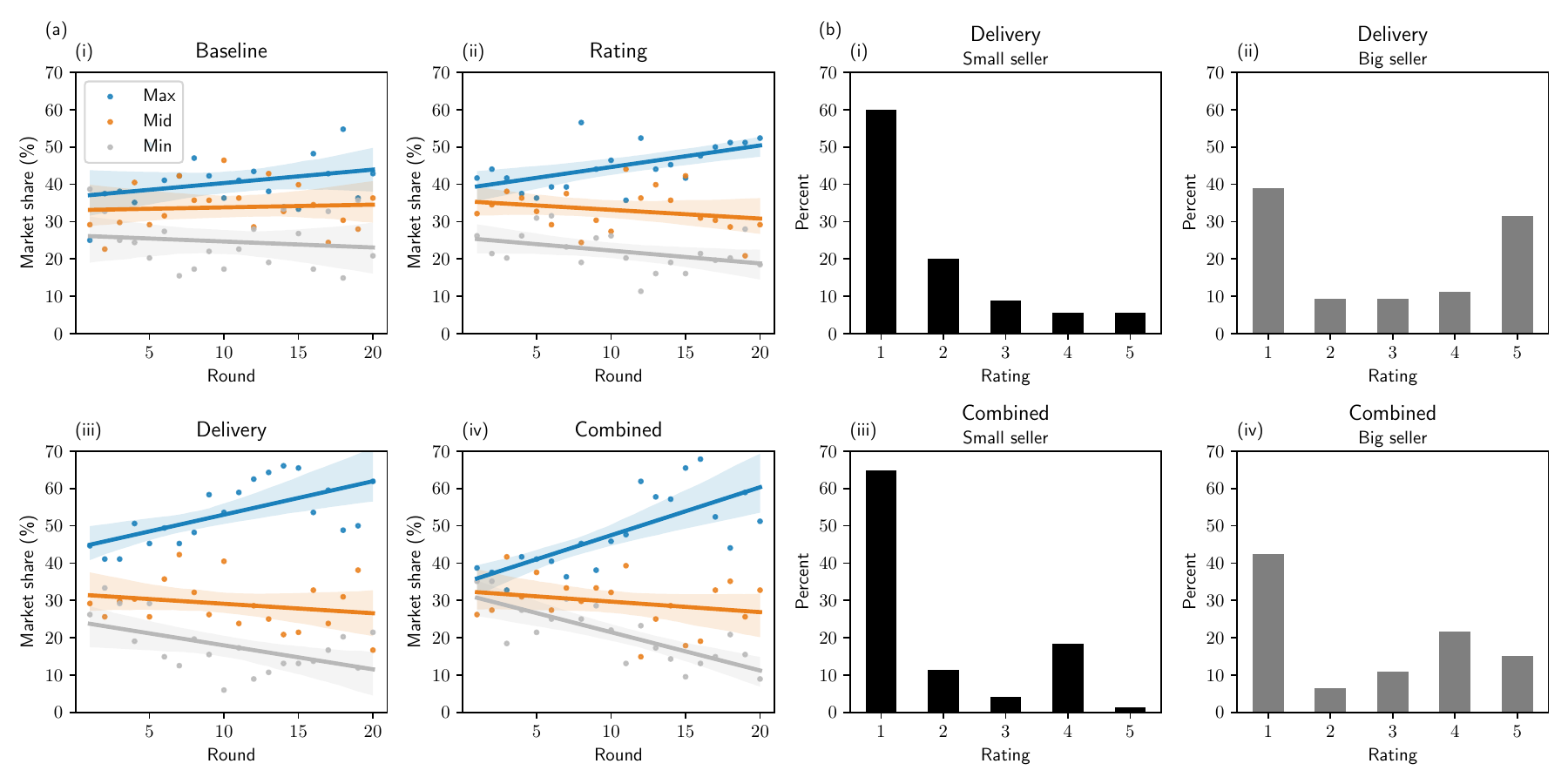}
    }
    \caption{The change in each seller's market share over time. \emph{\footnotesize (a) We plot the market share of the largest seller (blue), the intermediate seller (orange) and the smallest seller (grey) over time. Sellers are categorised according to their average market share over all market rounds. (b) We plot the frequency of submitted ratings for sellers affected by the delivery attack. We differentiate between sellers with two sales (big sellers) and sellers with one sale (small sellers). }}\label{fig::concentration_round}
    
\end{figure}

\noindent Aside from its impact on sellers' earnings, the delivery intervention changes the structure of the market. The panels in \cref{fig::concentration_round}{\color{blue}a} illustrate the evolution of market share held by the largest (blue), intermediate (orange) and smallest (grey) sellers over time divided by treatment. In the delivery treatments, the market share of the largest seller increases over time (W, $p<0.05$ for D and $p<0.001$ for C), while it stays unchanged in both the baseline and rating treatments. The largest seller's market share in the delivery and combined treatments is, respectively, 59\% and 56\%, compared to 41\% for the baseline treatment (MW, $p<0.01$ for D and C). 

The delivery intervention has an asymmetric effect on the reputation of sellers with two sales (big sellers) and sellers with one sale (small sellers). In each round, big sellers are less likely to have all their sales affected by a delivery attack than small sellers, with respective probabilities of 4\% and 20\%. \cref{fig::concentration_round}{\color{blue}b} plots the frequency of submitted ratings for big sellers and small sellers affected by a delivery attack. For big sellers, the positive reputational effects from successful delivery of one of their goods may counterbalance the negative reputational effects from the delivery attack. To investigate this mechanism, we distinguish between sellers for whom all sales are affected by the attack (complete delivery attack) and sellers partially affected by the attack (incomplete delivery attack) (see Appendix). The delivery attack increases failure to sell for completely affected sellers (\Cref{finding2tab1}, $p<0.001$) and has no effect on partially affected sellers (\Cref{finding2tab1}, $p=0.89$). This finding is robust to distinguishing between big sellers (\Cref{finding2tab1}, $p<0.01$) and small sellers (\Cref{finding2tab1}, $p<0.001$) completely affected by the delivery attack.

\section{\label{sec::conclusions} Discussion and Conclusions}

This paper has examined strategies to reduce the efficiency of illicit markets, inspired by the growth of cybercrime markets. These markets generate both negative economic and social effects, as well as challenges for traditional law enforcement. Alongside arrests, site takedowns and other conventional strategies, this paper has proposed a complementary strategy: to directly and indirectly attack the reputation of sellers, making the markets less efficient. As these harmful criminal markets are characterised by asymmetric information, we have focussed our analysis on how we could use reputational interventions to disrupt them. We show that the delivery intervention lowers market efficiency respectively by 70\% (delivery) and 76\% (combined) relative to the baseline, while the rating intervention has no effect.

An interesting consequence of the delivery intervention is that it affects the composition of the market differently than a rating intervention. In the delivery treatments, the market share of the dominant seller increases over time. From a policy point of view, this suggests that the intervention might lead to the formation of less competitive markets with power concentrated in a few hands. It is arguable whether this is a good or bad outcome. While it might lead to increasingly more powerful criminal actors, this process also creates focal points for law enforcement investigations. Authorities will be able to monitor larger providers than smaller ones and may decide that less harmful goods can be allowed on the site. 

The role of personal trading relationships may explain the ineffectiveness of the rating intervention. A buyer who repeatedly purchases from the same seller has little need to access truthful rating information and can instead rely on their personal trading experience. We find that personal trading relationships are important determinants of trade in all treatments. Buyers are more likely to purchase a good when their trading partner in the previous round has goods available to purchase (W, $p<0.01$ for B and $p<0.001$ for R, D and C).

Directions for future work include investigating alternative forms of the rating attack. For instance, only targeting the highest rated sellers with a rating attack that downgrades reputation might be both cheaper to execute and have a more profound impact on the market. Interventions that impact qualitative feedback, rather than numerical ratings, might also be considered. Variations of seller side attacks that differ from the particular delivery attack we tested, such as internal “undercover” variants, might also be examined. Our hope is that this study opens up a novel line of research to investigate economic disruptions to real world cybercrime and other illicit markets. More applied work should follow to refine the most effective interventions, matched to the empirical realities of specific illicit markets.

In opening this line of research, this study also raises an important ethical question: should we allow illegal markets to continue operating? The answer may seem straightforward when shutting down a market is the most effective intervention. The issue becomes more complex when the kinds of targeted disruptive interventions discussed in this paper are chosen alone. In a given case, this may improve overall societal welfare, but may also implicitly allow criminal offending to continue. As these kinds of interventions are explored further, this is a question that warrants careful consideration.

\clearpage
\renewcommand*{\bibfont}{\footnotesize}
\printbibliography[heading=bibintoc]

\clearpage

\appendix

\setcounter{figure}{0}
\setcounter{table}{0}

\renewcommand{\thefigure}{\thesection.\arabic{figure}}
\renewcommand{\thetable}{\thesection.\arabic{table}}

\section*{Appendices}
\section{Supplementary Text}

We begin by outlining our empirical strategy for the statistical analyses presented in the main text and Supplementary Materials. We then present summary statistics for our outcome variables, disaggregated by treatment group. Next, we conduct robustness checks to verify that our findings hold under parametric estimation methods. We then provide supplementary analyses: first, examining how the delivery intervention reduces market efficiency; and second, how it increases market concentration. Finally, we discuss the apparent ineffectiveness of the rating intervention.

\subsection{Empirical Strategy}

In the main text, we use a conservative, non-parametric testing procedure based on group averages\footnote{We compute the group average by weighting each round equally.} to address dependence between observations. Specifically, we test for treatment differences using the Mann-Whitney U test \parencite{Mann1947}, using one observation per group. Similarly, we test for time trends by comparing the group average over the first ten market rounds to the group average over the final ten market rounds, using Wilcoxon's signed-rank test \parencite{Wilcoxon1945}. Hereafter, we denote the Mann-Whitney U test by MW and Wilcoxon's signed-rank test by W. Due to significant time trends and our focus on equilibrium behaviour, we limit treatment comparisons in the main text to the final ten market rounds.

To verify that our results are robust to the choice of testing procedure, we estimate a series of random effect models for the key outcome variables discussed in the main text: market efficiency, seller earnings, buyer earnings, number of goods sold, buyer inactivity, and the market share of the largest seller. The random effects specification accounts for correlation in the error term at either the group or individual level, depending on the outcome (refer to \cref{fig::robustness_randomeffects}). Motivated by significant time trends and following the approach of \textcite{Cason2002}, we distinguish between short-run and long-run treatment effects (see \cref{eq::treatmenteffects}).
\begin{equation}\label{eq::treatmenteffects}
        y_{it}=\alpha_{SR} \left(\frac{1}{t}\right) + \alpha_{LR} \left(\frac{t-1}{t}\right)+\beta_{SR}\left( \frac{1}{t} \times T_{i} \right) + \beta_{LR} \left( \frac{t-1}{t} \times T_{i} \right) + u_{i} + \epsilon_{it}
\end{equation}Here, $y_{it}$ denotes the outcome variable, and $T_{i}$ is a vector of treatment indicators for the delivery, rating and combined treatment conditions. $t$ denotes the round number\footnote{The regression equation is estimated using data from all market rounds, as time trends and dependencies are accounted for parametrically.}. $u_{i}$ is a group- or individual-level random effect, and $\epsilon_{it}$ is an idiosyncratic error term. This specification allows for a time-varying treatment effect, where $\beta_{SR}$ captures the short-run effect and $\beta_{LR}$ captures the long-run effect.

To understand the effect of the delivery intervention on market concentration, we estimate a random effects probit model of the likelihood a seller fails to sell at least one of their goods on the delivery attack (see \cref{eq::RPE,eq::data-gen}).
\begin{align}\label{eq::RPE}
    y_{it} = 
    \begin{cases} 
    1 & \text{if } f_{it}^{*} > 0, \\
    0 & \text{otherwise}.
    \end{cases}
\end{align}
\begin{align}\label{eq::data-gen}
    f_{it}^{*} = \alpha_{0} + \alpha_{1}d_{it-1} + \alpha_{2} X_{it} + u_{i} + \epsilon_{it}
\end{align}
Here, $f_{it}$ equals unity if seller i fails to sell a good in round $t$, and $f_{it}^{*}$ is the data generating process. $\alpha_{i}$ is an individual-level random effect, and $\epsilon_{it}$ is an idiosyncratic error term. $D_{it-1}$ is equal to unity if seller i was affected by the delivery attack in the previous market round. $Z_{it}$ is a vector of controls, which includes the number of goods that seller $i$ sold in the previous market round\footnote{We include this to account for the fact that the likelihood of being affected by the delivery attack varies with the number of sales.}, the seller's advertisement in round $t$, demographic controls and round fixed effects. To aid the identification of mechanisms, we differentiate between sellers for whom all sales are affected by the delivery attack (complete delivery attack) and sellers for whom only some of their sales are affected by the delivery attack (incomplete delivery attack). In addition, we differentiate between sellers who sold two goods in the previous market round (big sellers) and sellers who sold one good (small sellers). Importantly, we restrict attention to the delivery treatment and the combined treatment where the delivery attack is active, and on sellers who were eligible for the delivery attack in the previous market round and have advertised goods in the current round. We also exclude the first market round, since sellers could not have been affected by the delivery attack in a previous round.

Finally, to understand the ineffectiveness of the rating intervention, we investigate the role of personal trading relationship in purchasing behaviour. We define a buyer's trading partner as the seller the buyer purchased from in the previous market round. We begin by comparing a buyer's likelihood of purchasing a good when their trading partner has a good available to purchase to a buyer's likelihood of purchasing a good when their trading partner does not have a good available to purchase, using W \parencite{Wilcoxon1945}. Secondly, we compare the observed likelihood that buyers purchase from their personal trading partner to the predicted likelihood under random choice, using W\footnote{For this measure, we restrict our attention to buyers who chose to purchase a good.}  \parencite{Wilcoxon1945}. To control for confounders, we estimate McFadden's choice model \parencite{McFadden1974} for each treatment condition using the STATA cmclogit command. We model a buyer's expected utility from purchasing a good as a function of the advertised quality, price, the seller's public reputation and whether the seller is the buyer's personal trading partner (see \cref{eq::cmclogit_good}).
\begin{align}\label{eq::cmclogit_good}
    U_{it}(g) = \alpha_{0} + \alpha_{1}\text{Type}_{jt} + \alpha_{2} \text{Price}_{jt} + \alpha_{3} \text{Reputation}_{jt} + \alpha_{4} \text{Partner}_{ijt} + \epsilon_{itg}
\end{align}
where $i$ denotes the buyer, $j$ denotes the seller and $t$ denotes the market round. We model a buyer's utility from not purchasing a good as a function of their benefit from certainty and their cognitive saving (see \cref{eq::cmclogit_nothing}).
\begin{align}\label{eq::cmclogit_nothing}
    V_{it}^{*} = \beta_{0} + \beta_{1} Z_{i} + \beta_{2} \text{Order}_{it} + \beta_{3} t + e_{it}
\end{align} 
where $i$ denotes the buyer and t denotes the market round. We proxy a buyer's benefit from certainty by their demographics ($Z_i$). We model the cognitive saving as a function of the buyer's purchase order ($\text{Order}_{it}$) and the market round. For the regression analysis, we focus on the final ten market rounds to mitigate missing rating issues and consider only buyers for whom a good is available to purchase. Additionally, we drop observations who timed out on the decision page.

\subsection{Summary Statistics}\label[appendix]{app::time_trends}

\Cref{tab::summarystat_outcome} presents summary statistics for our outcome variables separated by treatment. We define outcome variables that we have constructed in \cref{tab::definitions}. As shown in \cref{tab::summarystat_outcome}, there are significant time trends in several of our outcome variables, including seller earnings (W, $p<0.01$ for D and C), the number of goods sold (W, $p<0.05$ for D and $p<0.01$ for C) and the dominant seller's market share (W, $p<0.05$ for D and $p<0.001$ for C). 

{
    \footnotesize
    \centering
    \begin{longtable}{l*{5}{c}}
        \caption{Summary statistics for outcome variables}\label{tab::summarystat_outcome} \tabularnewline
        \toprule
        Variable & Rounds          &          Baseline &           Rating &           Delivery &           Combined \\
        \hline
        Market efficiency & 1-10 &       34.47&       27.71&       16.51$^{*}$ &       10.47$^{***}$ \\
             &       &     (5.282)&     (4.103)&     (3.951)&     (2.145)\\
        & 11-20 &       29.14 &       34.92 &       8.673$^{**}$ &       6.939$^{**}$ \\
                 &   &     (4.456)&     (6.000)&     (6.108)&     (4.144)\\
        & 1-20 &       31.81&       31.32&       12.59$^{**}$&       8.704$^{***}$\\
               &     &     (4.659)&     (4.582)&     (4.429)&     (2.274)\\
               & W & $p\geq0.1$ & $p\geq0.1$ & $p\geq0.1$ & $p\geq0.1$ \\
        [1em]
        Adjusted market efficiency & 1-10 &       34.47&       27.71&       25.12&       15.93$^{*}$\\
             &       &     (5.282)&     (4.103)&     (6.012)&     (3.265)\\
        & 11-20 &       29.14&       34.92&       13.20$^{*}$&       10.56$^{*}$\\
                &    &     (4.456)&     (6.000)&     (9.295)&     (6.305)\\
        & 1-20 &       31.81&       31.32&       19.16&       13.25$^{**}$\\
                    &    & (4.659)&     (4.582)&     (6.739)&     (3.461)\\
                    & W & $p\geq0.1$ & $p\geq0.1$ & $p\geq0.1$ & $p\geq0.1$ \\
        [1em]
        Seller earnings & 1-10 &       48.99&       40.67&       33.46&       34.98\\
                    &  &   (6.566)&     (5.173)&     (7.662)&     (6.076)\\
        & 11-20 &       49.61&       38.48&       18.40$^{**}$&       21.10$^{***}$\\
                    &   &  (5.796)&     (5.953)&     (7.264)&     (4.173)\\
        & 1-20 &       49.30&       39.57&       25.93$^{**}$&       28.04$^{*}$\\
                    &   &  (5.534)&     (5.176)&     (7.204)&     (4.470)\\ 
             & W & $p\geq0.1$ & $p\geq0.1$ & $p<0.01$ & $p<0.01$ \\        [1em]
        Buyer earnings & 1-10 &      -6.579&      -6.252&      -10.65&      -17.08\\
                    &   &  (7.779)&     (4.642)&     (5.805)&     (4.599)\\
        & 11-20 &      -11.71&       1.695&      -6.209&      -9.755\\
                    & &    (5.139)&     (4.960)&     (6.559)&     (3.908)\\
        & 1-20 &      -9.142&      -2.279&      -8.429&      -13.42\\
                    &  &   (5.971)&     (4.361)&     (5.984)&     (4.013)\\
        & W & $p\geq0.1$ & $p\geq0.1$ & $p\geq0.1$ & $p<0.05$ \\  [1em]
        
        Number of goods sold & 1-10 &       2.943&       2.614&       2.707&       2.779\\
                    & &    (0.120)&     (0.126)&     (0.122)&     (0.130)\\
        & 11-20 &       2.836&       2.593&       2.329$^{*}$&       2.329$^{**}$\\
                    &  &   (0.120)&     (0.178)&     (0.173)&     (0.144)\\
        & 1-20 &       2.889&       2.604&       2.518&       2.554\\
                    &  &   (0.103)&     (0.141)&     (0.134)&     (0.121)\\
                    & W & $p\geq0.1$ & $p\geq0.1$ & $p<0.05$ & $p<0.01$ \\ [1em]

        Buyer inactivity rate (\%) & 1-10 &       18.27&       31.37$^{**}$&       29.40$^{*}$&       26.71\\
                    &   &  (2.582)&     (3.468)&     (3.451)&     (3.047)\\
        & 11-20 &       24.64&       34.11&          40.00$^{*}$&       40.95$^{**}$\\
                    &    & (2.952)&     (4.596)&     (4.889)&     (3.608)\\
        & 1-20 &       21.51&       32.74$^{*}$&       34.70$^{*}$&       33.84$^{**}$\\
                    &    & (2.577)&     (3.811)&     (3.875)&     (2.989)\\
        & W & $p<0.01$ & $p\geq 0.1$ & $p<0.05$ & $p<0.001$ \\[1em]

        Dominant seller's market   & 1-10 &       39.52&       42.68&       47.74&       39.76\\
        share (\%)            &    & (2.194)&     (2.664)&     (4.274)&     (2.504)\\
        & 11-20  &       41.49&       47.14&       59.11$^{**}$&       56.43$^{**}$ \\
                    &   &  (1.845)&     (3.797)&     (4.913)&     (3.260)\\
        & 1-20  &       40.51&       44.91&       53.42$^{**}$ &       48.10$^{*}$\\
                    &   &  (1.564)&     (2.275)&     (4.243)&     (2.527)\\
        & W & $p\geq0.1$ & $p\geq0.1$ & $p<0.05$ & $p<0.001$ \\ [1em]  

        Herfindahl-Hirschman index & 1-10 &      4564.5&      5276.8&      5397.8&      5120.0\\
        & &    (212.5)&     (292.4)&     (366.4)&     (166.9)\\
& 11-20 &      4693.5&      5224.2&      6245.0$^{**}$&      6049.6$^{***}$\\
        &   &  (213.9)&     (297.7)&     (462.7)&     (299.1)\\
& 1-20 &      4629.0&      5250.5&      5821.4$^{*}$ &      5584.8$^{**}$\\
        &    & (198.6)&     (260.4)&     (391.4)&     (191.2)\\
& W & $p\geq0.1$ & $p\geq0.1$ & $p<0.01$ & $p<0.001$ \\[1em]

        Purchased good:   & 1-10 & 91.00 & 82.38$^{*}$ & 79.53$^{*}$ & 81.09$^{*}$ \\
        \emph{partner available} (\%) & & (1.915) & (2.697) & (4.265) & (3.994) \\
        & 11-20 & 87.57 & 88.84 & 80.69 & 79.64 \\
        & & (2.360) & (2.142) & (4.069) & (3.203)  \\
        & 1-20 & 89.13 & 85.93 & 80.36 & 80.56$^{*}$ \\
        & & (1.608) & (1.784) & (3.391) & (2.872)  \\
        & W & $p>0.10$ & $p<0.10$ & $p>0.10$ & $p>0.10$\\ [1em]
        
        Purchased good:   & 1-10 & 71.45 & 61.01 & 51.22$^{*}$ & 73.13 \\
        \emph{partner unavailable} (\%) & & (6.047) & (4.239) & (7.326) & (3.821) \\
        & 11-20 & 54.38 & 35.22 & 40.57 & 55.08 \\
        & & (7.642) & (6.967) & (7.729) & (3.966) \\
        & 1-20 & 63.31 & 49.88 & 47.72 & 64.68 \\
        & & (5.593) & (4.548) & (6.627) & (2.000) \\
        & W & $p<0.05$ & $p<0.01$ & $p>0.10$ & $p<0.01$ \\ [1em]
        
        Trading partner (\%) & 1-10 & 39.09 & 31.72 & 41.71 & 37.20 \\
        & & (4.407) & (3.761) & (4.726) & (3.679) \\
        & 11-20 & 53.38 & 49.40 & 62.14 & 47.45 \\
        & & (4.592) & (4.055) & (4.704) & (3.494) \\
        & 1-20 & 46.77 & 40.47 & 52.05 & 42.71 \\
        & & (4.145) & (2.989) & (3.395) & (1.859) \\
        & W & $p<0.01$ & $p<0.01$ & $p<0.01$ & $p>0.10$\\
        
        \bottomrule
        
    \end{longtable}
    {Notes: The table contains sample means with standard errors in parentheses. The starred p-values are derived from Mann-Whitney U tests comparing each treatment to the baseline treatment, where $^{*} \; p<0.05, ^{**} \; p<0.01, ^{***} \; p<0.001$. We report p-values from Wilcoxon's signed-rank tests for time trends in rows labelled $W$.}
}

\begin{table}[bhpt] 
    \caption{Definitions of constructed variables}\label{tab::definitions}
    \centering 
    \begin{threeparttable}
    \begin{tabular}{p{0.4\textwidth} p{0.5\textwidth}}
        
        \toprule
        Variable & Definition  \\
        \hline
         Market efficiency & Sum of player earnings normalised by the maximum gains from trade in the Baseline treatment$^{1}$ \\ 
         Adjusted market efficiency & Sum of player earnings normalised by the maximum expected gains from trade in the Delivery treatment$^{2}$  \\
         Buyer inactivity rate  & The proportion of active$^{3}$ buyers who choose to not purchase a good given a good is available to purchase.  \\
         Dominant seller's market share & The dominant seller is the seller with the largest average market share over all trading periods \\
         Purchased good: \emph{partner available}  & The proportion of active$^{3}$ buyers who choose to purchase a good given their personal trading partner has a good available to purchase \\
         Purchased good: \emph{partner unavailable} & The proportion of active$^{3}$ buyers who choose to purchase a good given their personal trading partner does not have a good available to purchase. We exclude observations for which there are no goods available to purchase. \\
         Trading partner & The proportion of purchasing buyers that choose to purchase from their personal trading partner given their personal trading partner has a good available to purchase   \\
         \bottomrule
    \end{tabular}
    
    \begin{tablenotes}
    \small
    \item $^{1}$ The maximum gains from trade are realised when only high (super) quality goods are traded. Each high (super) quality good is valued at 150 points, resulting in a total valuation of 600 points. Production cost is minimised when two sellers each produce one high (super) quality good and one seller produces two high (super) quality goods, resulting in a total cost of 250 points. Hence, the maximum surplus is 600 points - 250 points = 350 points. $^{2}$ Similarly, the maximum gains from trade are realised when only high (super) quality goods are traded. There is a 20\% likelihood that each good does not arrive. Hence, the total (expected) valuation is 480 points. The total cost is unchanged at 250 points. Hence, the (expected) maximum surplus is 480 points - 250 points = 230 points. $^{3}$ A buyer is active if they do not time out on the buyer decision page. 
\end{tablenotes}
\end{threeparttable}
\end{table} 

\subsection{Robustness Checks}

\Cref{fig::robustness_randomeffects} reports maximum likelihood estimates from random effects models for our key outcomes: market efficiency (col. 1), seller earnings (col. 2), buyer earnings (col. 3), the number of goods sold (col. 4), buyer inactivity (col. 5), and the market share of the largest seller (col. 6)\footnote{The respective units of these measures are: percent (market efficiency), experimental currency units (seller earnings), experimental currency units (buyer earnings), goods (number of goods sold), percent (buyer inactivity) and percent (market share).}. 

Our first main result - that the delivery intervention reduces market efficiency - is robust to the choice of testing procedure. Market efficiency in the delivery and combined treatments is, respectively, 60\% and 77\% lower than in the baseline in the long-run ($p<0.01$ for D, and $p<0.001$ for C). This impact is concentrated among sellers. Sellers' earnings in the delivery and combined treatments are, respectively, 55\% and 45\% lower than in the baseline in the long-run ($p<0.01$ for C and D). In contrast, the rating intervention has an insignificant effect on both market efficiency and seller earnings. Additionally, buyer earnings are unaffected by the treatment.

Our second main result - that the delivery intervention reduces the number of goods sold - is robust to this testing procedure. The delivery and combined treatments reduce the number of goods sold by roughly half a unit relative to the baseline in the long run, corresponding to an approximate 17\% reduction in market activity ($p < 0.05$ for D and $p < 0.01$ for C). Moreover, both the delivery and combined treatments exhibit a significant downward trend in the number of goods sold  ($p < 0.05$ for D, and $p < 0.001$ for C). This pattern is reflected on the demand-side. The proportion of active  buyers who choose not to purchase is roughly 60\% higher in the delivery and combined treatments than in the baseline in the long run ($p < 0.01$ for D, and $p < 0.001$ for C). Similarly, both the delivery and combined treatments exhibit a significant upward trend in the rate of buyer inactivity ($p < 0.01$ for D, and $p < 0.001$ for C). In contrast to our parametric results, the rate of buyer inactivity is significantly higher in the rating treatment than in the baseline in the long run ($p < 0.05$ for R). These findings should be interpreted with caution, however, as they are not robust to our parametric specifications.

Finally, our third main finding - that the delivery intervention increases market concentration - is also robust to this testing procedure. The market share of the largest seller is 31\% and 18\% higher in the delivery and combined treatments, respectively, than in the baseline ($p < 0.01$ for D, and $p < 0.05$ for C). In contrast, the rating intervention does not have a significant effect on market concentration.

\subsection{The Delivery Intervention Reduces Market Efficiency}

The delivery intervention has both a direct effect and an indirect effect on market efficiency. The delivery intervention reduces buyer surplus through good seizures (direct) and reduces the number of goods sold (indirect). To assess the indirect effect, we define adjusted market efficiency which adjusts the maximum gains from trade for the direct effect of good seizures (see \Cref{tab::summarystat_outcome,tab::definitions}). Adjusted market efficiency in the delivery and combined treatments is, respectively, 55\% and 64\% lower than in the baseline over the last 10 market rounds (MW, $p< 0.05$ for D and C). Thus, the decrease in market efficiency remains significant after adjusting the maximum gains from trade for

\begin{landscape}

\begin{table}[htbp]
    
    \centering 

\begin{threeparttable}

\def\sym#1{\ifmmode^{#1}\else\(^{#1}\)\fi}

\caption{MLE of Random Effects Model for Key Outcome Variables \label{fig::robustness_randomeffects}}
\scriptsize 
\begin{tabular}{l c c c c c c }
\toprule
                    &\multicolumn{1}{c}{(1)}&\multicolumn{1}{c}{(2)}&\multicolumn{1}{c}{(3)}&\multicolumn{1}{c}{(4)}&\multicolumn{1}{c}{(5)}&\multicolumn{1}{c}{(6)}\\
                    &\multicolumn{1}{c}{market efficiency}&\multicolumn{1}{c}{seller earnings}&\multicolumn{1}{c}{buyer earnings}&\multicolumn{1}{c}{number of goods sold}&\multicolumn{1}{c}{buyer inactivity}&\multicolumn{1}{c}{market share}\\
\midrule
$\frac{1}{t}$           &       39.19\sym{***}&       49.62\sym{***}&      -2.927         &       2.810\sym{***}&       13.77\sym{*}  &       26.91\sym{***}\\
                    &     (7.440)         &     (10.66)         &     (10.40)         &     (0.330)         &     (5.581)         &     (5.136)         \\
\addlinespace
$\frac{t-1}{t}$            &       30.19\sym{***}&       49.22\sym{***}&      -10.51\sym{*}  &       2.907\sym{***}&       24.51\sym{***}&       43.49\sym{***}\\
                    &     (4.240)         &     (5.477)         &     (5.230)         &     (0.117)         &     (2.870)         &     (1.789)         \\
\addlinespace
rating $\times$ $\frac{1}{t}$ &      -10.81         &      -5.423         &      -5.387         &       0.152         &       8.010         &       11.51         \\
                    &     (8.062)         &     (14.11)         &     (12.21)         &     (0.385)         &     (7.761)         &     (8.143)         \\
\addlinespace
rating $\times$ $\frac{t-1}{t}$ &       1.773         &      -10.67         &       9.550         &      -0.382         &       11.46\sym{*}  &       2.846         \\
                    &     (7.174)         &     (8.225)         &     (6.916)         &     (0.205)         &     (5.261)         &     (3.123)         \\
\addlinespace
delivery$\times$ $\frac{1}{t}$ &      -24.54\sym{**} &      -5.631         &      -17.25         &       0.297         &       3.566         &       9.930         \\
                    &     (9.322)         &     (14.46)         &     (13.45)         &     (0.408)         &     (8.237)         &     (7.965)         \\
\addlinespace
delivery $\times$ $\frac{t-1}{t}$ &      -18.05\sym{**} &      -27.26\sym{**} &       4.654         &      -0.518\sym{*}  &       14.71\sym{**} &       13.57\sym{**} \\
                    &     (6.593)         &     (9.366)         &     (8.218)         &     (0.203)         &     (5.197)         &     (4.893)         \\
\addlinespace
combined $\times$ $\frac{1}{t}$ &      -23.01\sym{*}  &      -17.64         &      -6.899         &       0.451         &       0.370         &       2.312         \\
                    &     (9.540)         &     (13.12)         &     (11.93)         &     (0.387)         &     (7.861)         &     (7.434)         \\
\addlinespace
combined $\times$ $\frac{t-1}{t}$ &      -23.12\sym{***}&      -22.05\sym{**} &      -3.698         &      -0.508\sym{**} &       14.89\sym{***}&       8.747\sym{*}  \\
                    &     (5.304)         &     (7.304)         &     (6.762)         &     (0.175)         &     (4.394)         &     (3.421)         \\
\midrule
$\sigma_{u}$             &                     &                     &                     &                     &                     &                     \\
constant            &       13.66\sym{***}&       29.98\sym{***}&       23.04\sym{***}&       0.424\sym{***}&       22.64\sym{***}&       8.696\sym{***}\\
                    &     (1.484)         &     (3.079)         &     (2.440)         &    (0.0426)         &     (1.158)         &     (1.661)         \\
\midrule
$\sigma_{e}$            &                     &                     &                     &                     &                     &                     \\
constant            &       25.55\sym{***}&       63.39\sym{***}&       49.56\sym{***}&       0.708\sym{***}&       40.16\sym{***}&       23.84\sym{***}\\
                    &     (0.811)         &     (2.127)         &     (1.318)         &    (0.0180)         &     (0.776)         &     (0.962)         \\
\midrule
random effect & \text{group} & \text{individual} & \text{individual} & \text{group} & \text{individual} & \text{group} \\
market rounds & 1-20 & 1-20 & 1-20 & 1-20 & 1-20 & 1-20 \\
\midrule
no. observations        &        1120         &        3360         &        4480         &        1120         &        4325         &        1120         \\
\bottomrule
\end{tabular}

\begin{tablenotes}
      \item Notes: $^{*} p<0.05$, $^{**} p<0.01$ and $^{***} p<0.001$. Cluster robust standard errors in parentheses by group.
\end{tablenotes}

\end{threeparttable}

\end{table}

\end{landscape}

\noindent the direct effect of good seizures.

\subsection{The Delivery Intervention Increases Market Concentration}

The effect of the delivery intervention on market concentration is robust to our measure of market concentration. The dominant seller's market share in the delivery and combined treatments is, respectively, 42\% and 36\% higher than in the baseline over the last 10 market rounds (MW, $p<0.01$ for D and C). Similarly, the Herfindahl-Hirschman index in the delivery and combined treatments is, respectively, 33\% and 29\% higher than in the baseline over the last 10 market rounds (MW, $p<0.01$ for D and $p<0.001$ for C).

In the main text, we discuss that the delivery intervention may lead to an increase in market concentration, due to its asymmetric effect on the reputation of small sellers and big sellers. In each round, big sellers are less likely to have all their sales affected by a delivery attack than small sellers, with respective probabilities of 4\% and 20\%. To test this line of reasoning, we investigate how the delivery attack affects the likelihood a seller fails to sell at least one of their goods, differentiating between complete delivery attacks and incomplete delivery attacks (see \cref{finding2tab1}).

We find that the delivery attack increases the likelihood that a seller fails to sell at least one of their goods in the next market round (col. 1, $p<0.001$), which is robust to controlling for demographics and quiz performance (col. 2, $p<0.001$). Upon further inspection, we find that the complete delivery attack increases the likelihood that a seller does not sell all of their goods (col. 3, $p<0.001$) but the incomplete delivery attack has no effect (col. 3, $p=0.89$). Moreover, the complete delivery attack has a significantly larger effect than the incomplete delivery attack ($p<0.01$). This is consistent with the delivery intervention increasing market concentration through its asymmetric effect on the reputation of small sellers and big sellers.

One concern is that we are picking up heterogeneity by firm size rather than type of delivery attack per se. To address this concern, we differentiate between complete delivery attacks affecting small sellers and big sellers. We find that the complete delivery attack increases the likelihood that a seller fails to sell at least one of their goods for both small sellers (col. 4, $p<0.001$) and big sellers (col. 4, $p<0.01$). This suggests that firm size is not the driver of our result.

\subsection{Personal Trading Relationships}

One possible mechanism behind the ineffectiveness of the rating attack is the role of personal trading relationships. In our experimental design, buyers can form individual relationships with sellers, as each seller is identified by a unique nickname. In this context, buyers do not need to rely on public reputation information, instead, they can use the private information accumulated through repeated interactions with specific sellers. While it is challenging to isolate the mediating effect

\begin{landscape}
    
\begin{table}[htbp]
    \centering
    
\def\sym#1{\ifmmode^{#1}\else\(^{#1}\)\fi}

\begin{threeparttable}
    
\caption{MLE of Random Effects Probit Model for Failure to Sell \label{finding2tab1}}
\scriptsize 
\begin{tabular}{l*{4}{D{.}{.}{-1}}}
\toprule
                    &\multicolumn{1}{c}{(1)}&\multicolumn{1}{c}{(2)}&\multicolumn{1}{c}{(3)}&\multicolumn{1}{c}{(4)}\\
                    &\multicolumn{1}{c}{failure to sell}&\multicolumn{1}{c}{failure to sell}&\multicolumn{1}{c}{failure to sell}&\multicolumn{1}{c}{failure to sell}\\
\midrule
L. delivery attack      &       0.457\sym{***}&       0.457\sym{***}&                     &                     \\
                    &     (0.119)         &     (0.119)         &                     &                     \\

\addlinespace
L. complete delivery attack    &                     &                     &       0.686\sym{***}&                     \\
                    &                     &                     &     (0.145)         &                     \\
\addlinespace
L. incomplete delivery attack     &                     &                     &     -0.0299         &      0.0383         \\
                    &                     &                     &     (0.208)         &     (0.211)         \\
\addlinespace
L. complete delivery attack: small seller &                     &                     &                     &       0.585\sym{***}\\
                    &                     &                     &                     &     (0.150)         \\
\addlinespace
L.complete delivery attack: big seller &                     &                     &                     &       2.228\sym{**} \\
                    &                     &                     &                     &     (0.723)         \\
\addlinespace
L. number of goods sold &      -0.384\sym{**} &      -0.388\sym{**} &      -0.202         &      -0.294         \\
                    &     (0.137)         &     (0.136)         &     (0.152)         &     (0.157)         \\
\addlinespace
number of goods advertised &       1.394\sym{***}&       1.417\sym{***}&       1.449\sym{***}&       1.467\sym{***}\\
                    &     (0.141)         &     (0.138)         &     (0.141)         &     (0.142)         \\
\addlinespace
super advertised &      -1.136\sym{***}&      -1.108\sym{***}&      -1.097\sym{***}&      -1.089\sym{***}\\
                    &     (0.147)         &     (0.147)         &     (0.148)         &     (0.149)         \\
\addlinespace
price         &      0.0127\sym{***}&      0.0126\sym{***}&      0.0127\sym{***}&      0.0125\sym{***}\\
                    &   (0.00166)         &   (0.00166)         &   (0.00168)         &   (0.00167)         \\
\addlinespace
constant            &      -2.917\sym{***}&      -2.637\sym{***}&      -2.965\sym{***}&      -2.870\sym{***}\\
                    &     (0.350)         &     (0.765)         &     (0.787)         &     (0.787)         \\
\midrule
lnsig2u             &      -0.839\sym{**} &      -1.412\sym{***}&      -1.359\sym{***}&      -1.378\sym{***}\\
                    &     (0.277)         &     (0.337)         &     (0.333)         &     (0.335)         \\
\midrule
round fixed effects & Y & Y & Y & Y \\
demographics & N & Y & Y & Y \\
market rounds & 2-20 & 2-20 & 2-20 & 2-20 \\
\midrule
no. observations        &        1069         &        1069         &        1069         &        1069         \\
\bottomrule
\end{tabular}

\begin{tablenotes}
    \scriptsize 
    \item Notes: Standard errors are clustered at the group level. Significance levels: \sym{*} \(p<0.05\), \sym{**} \(p<0.01\), \sym{***} \(p<0.001\). We control for the demographics listed in \cref{tab::summary_stats} as well as the number of quiz attempts. 
\end{tablenotes}

\end{threeparttable}

\end{table}
\end{landscape}

\noindent of these personal relationships in our analysis, since both reputation mechanisms are present in all treatments, we can nonetheless investigate whether personal trading relationships are an important determinant of trade.

Our results indicate that personal trading relationships are a key driver of trade across all treatments. Buyers are more likely to purchase a good when their trading partner in the previous round has goods available to purchase (W, $p<0.01$ for B and $p<0.001$ for R, D and C). Moreover, the observed likelihood that buyers purchase from their trading partner is greater than the predicted likelihood under random choice. This finding is robust to generating predictions based on the number of sellers ($p<0.01$ for B, R and C and $p<0.001$ for D) or the number of goods ($p<0.01$ for C, $p<0.05$ for B and R and $p< 0.001$ for D). A limitation of this non-parametric approach is that it cannot fully account for confounding factors, such as public reputation, that might also contribute to this effect. To control for confounders, we estimate McFadden's choice model for each treatment condition (see \Cref{tab::trading}). We find that buyers are more likely to purchase a good from their personal trading partner ($p<0.001$ for all), which is robust to controlling for advertisement properties and the seller's public reputation ($p<0.001$ for B, R and D and $p<0.01$ for C). This suggests that personal trading relationships play a significant role in shaping trade, which may reduce buyers' reliance on the private reputation system and thereby diminish the effectiveness of the rating attack.

\section{Materials and Methods}

We describe how the experiment was implemented, describe the dataset and present the experiment instructions.

\subsection{Data Collection}

We describe how subjects were recruited for the experiment and outline the workflow of the experiment.

\subsubsection{Recruitment}

We recruited participants from Amazon Mechanical Turk. We restricted our sample to workers resident in the USA with a HIT approval rating greater than 95\% and with more than 500 HITs approved. We impose approval requirements to improve data quality. High-reputation workers have been found to fail attention check questions less often and exhibit greater internal consistency in online experiments \parencite{Paolacci2010,Peer2013}. 

\begin{figure}[b!]
    
    \centering
      \begin{subfigure}[t]{0.3\textwidth}
      
        \includegraphics[width=\textwidth]{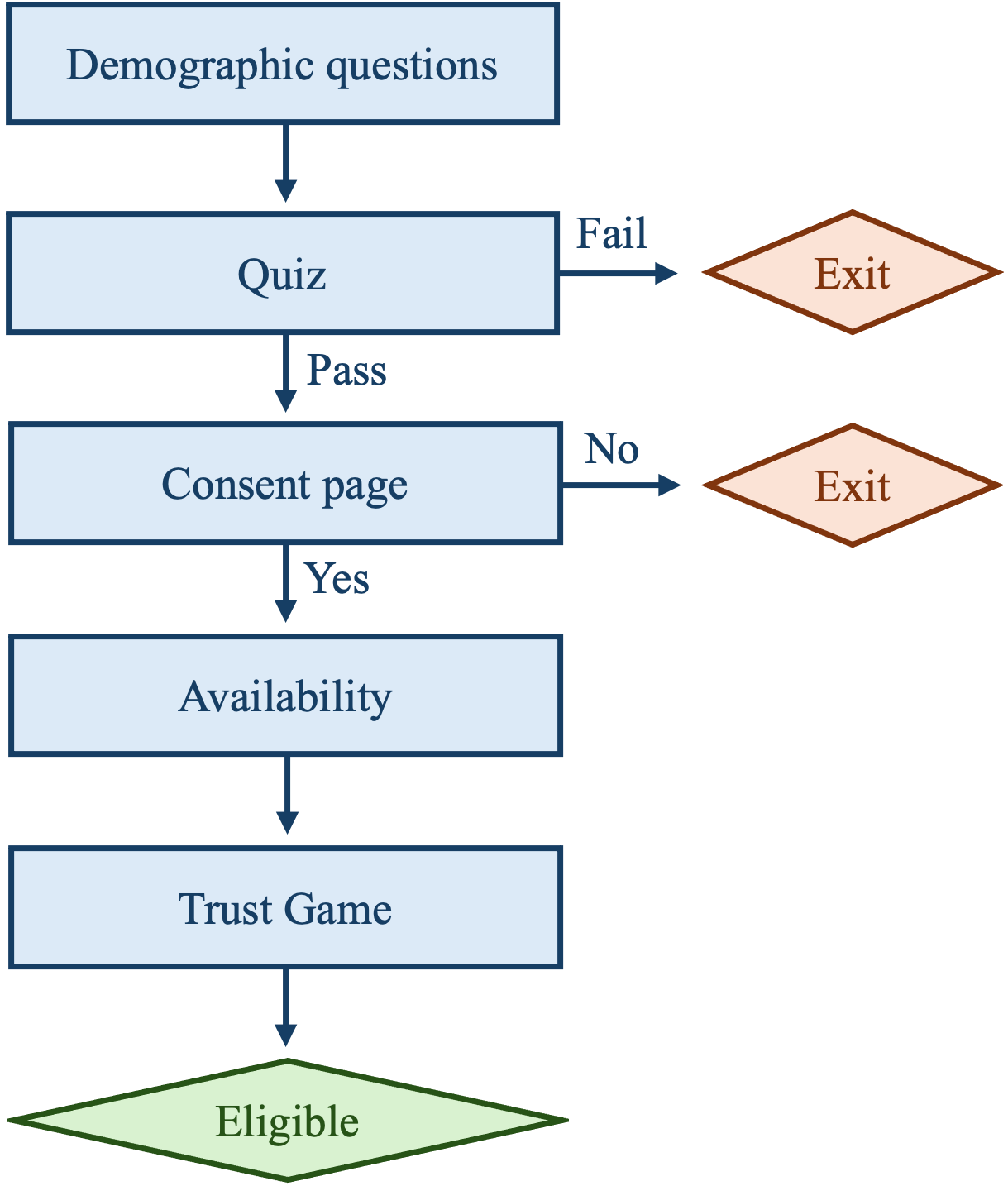}

        \caption{Recruitment survey}\label{fig::recruitment_survey}
        
      \end{subfigure}
      ~ \hspace{1em}
      \begin{subfigure}[t]{0.3\textwidth}
      
        \includegraphics[width=\textwidth]{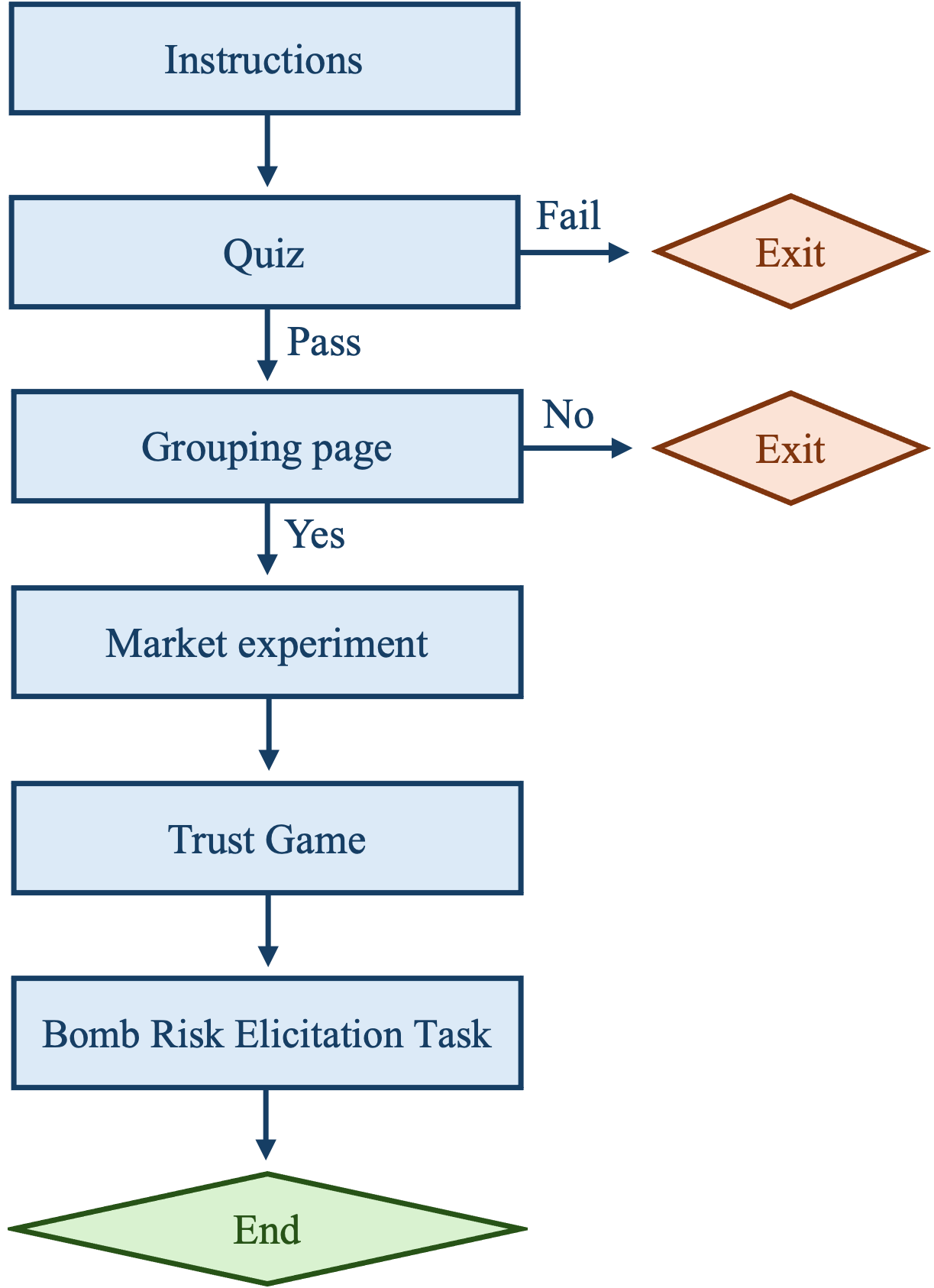}
        
        \caption{Experiment}\label{fig::experiment_workflow}

      \end{subfigure}

    \caption{Workflows}

    \end{figure}

Participants completed a short recruitment survey (median completion time of 3.58 minutes). The workflow of the recruitment survey is summarised in \cref{fig::recruitment_survey}. We paid participants \$0.50 

\begin{landscape}

\begin{table}[htbp]
    \centering
    \footnotesize 

    \begin{threeparttable}

\def\sym#1{\ifmmode^{#1}\else\(^{#1}\)\fi}
\caption{MLE of Conditional Logit Model of Buyer Choice \label{tab::trading}}
\begin{tabular}{l c c c c c c c c}
\toprule
                    &\multicolumn{1}{c}{(1)}&\multicolumn{1}{c}{(2)}&\multicolumn{1}{c}{(3)}&\multicolumn{1}{c}{(4)}&\multicolumn{1}{c}{(5)}&\multicolumn{1}{c}{(6)}&\multicolumn{1}{c}{(7)}&\multicolumn{1}{c}{(8)}\\
                    &\multicolumn{1}{c}{choice}&\multicolumn{1}{c}{choice}&\multicolumn{1}{c}{choice}&\multicolumn{1}{c}{choice}&\multicolumn{1}{c}{choice}&\multicolumn{1}{c}{choice}&\multicolumn{1}{c}{choice}&\multicolumn{1}{c}{choice}\\
\midrule
price               &    -0.00799\sym{**} &     -0.0106\sym{**} &    -0.00932\sym{***}&     -0.0135\sym{***}&    -0.00795\sym{*}  &     -0.0112\sym{***}&     -0.0137\sym{***}&     -0.0161\sym{***}\\
                    &   (0.00302)         &   (0.00324)         &   (0.00280)         &   (0.00387)         &   (0.00333)         &   (0.00291)         &   (0.00352)         &   (0.00373)         \\
\addlinespace
regular             &       1.049         &      -0.367         &       2.470         &      -1.731         &       0.664         &      -1.737         &     -0.0929         &      -3.263\sym{*}  \\
                    &     (1.080)         &     (1.213)         &     (1.290)         &     (1.532)         &     (1.016)         &     (1.152)         &     (1.293)         &     (1.390)         \\
\addlinespace
super               &       1.764         &       0.460         &       3.598\sym{**} &      -0.864         &       1.746         &      -0.881         &       0.857         &      -2.248         \\
                    &     (1.096)         &     (1.229)         &     (1.268)         &     (1.511)         &     (1.124)         &     (1.238)         &     (1.307)         &     (1.365)         \\
\addlinespace
trading partner &       0.822\sym{***}&       0.943\sym{***}&       1.126\sym{***}&       1.088\sym{***}&       1.295\sym{***}&       1.016\sym{***}&       0.900\sym{***}&       0.585\sym{**} \\
                    &     (0.185)         &     (0.217)         &     (0.191)         &     (0.195)         &     (0.199)         &     (0.190)         &     (0.179)         &     (0.199)         \\
\addlinespace
average rating      &                     &       0.773\sym{***}&                     &       1.406\sym{***}&                     &       0.929\sym{***}&                     &       0.928\sym{***}\\
                    &                     &     (0.176)         &                     &     (0.260)         &                     &     (0.175)         &                     &     (0.217)         \\
\addlinespace
average new rating         &                     &       0.309\sym{***}&                     &       0.212\sym{*}  &                     &       0.285\sym{***}&                     &       0.163         \\
                    &                     &    (0.0846)         &                     &    (0.0887)         &                     &    (0.0772)         &                     &    (0.0845)         \\
\addlinespace
no new ratings              &                     &       1.560\sym{***}&                     &       0.708\sym{*}  &                     &       0.404         &                     &      0.0404         \\
                    &                     &     (0.389)         &                     &     (0.332)         &                     &     (0.350)         &                     &     (0.344)         \\
\midrule
round fixed effects & Y & Y & Y & Y & Y & Y & Y & Y \\
demographics & Y & Y & Y & Y & Y & Y & Y & Y \\
market rounds & 11-20 & 11-20 & 11-20 & 11-20 & 11-20 & 11-20 & 11-20 & 11-20 \\
\midrule
treatment & baseline & baseline & rating & rating & delivery & delivery & combined & combined \\
no. observations        &        1672         &        1672         &        1785         &        1785         &        1822         &        1822         &        1873         &        1873         \\
\bottomrule
\end{tabular}

\begin{tablenotes} \scriptsize 
    \item Notes: Standard errors are clustered at the individual level. Significance levels: \sym{*} \(p<0.05\), \sym{**} \(p<0.01\), \sym{***} \(p<0.001\). We control for the demographics listed in \cref{tab::summary_stats} as well as the number of quiz attempts. \emph{average new rating} represents the mean of ratings submitted for the seller in the previous market round. \emph{no new ratings} equals unity if the seller received no ratings in the previous round.
\end{tablenotes}

\end{threeparttable}

\end{table}

\end{landscape}

\noindent upon successful completion of the survey.

The quiz comprised two basic numeracy questions, involving multiplication and subtraction, and a consistency check question. For the consistency check question, participants had to re-enter their age, which was validated against their response in the first section of the survey. Participants had three attempts to correctly answer the quiz questions. The quiz failure rate was 2.18\%.

At the end of the survey, participants played the role of sender in the Trust Game~\parencite{Berg1995}. Participants were endowed with \$1.00 and chose how much to send to an anonymous participant. Each participant could send any amount of this \$1.00 in \$0.10 intervals to their pair. Participants were only paid for this bonus task if they participated in the experiment, which was communicated to participants.

We recruited participants between 28 February 2022 and 13 April 2022. Our final subject pool comprised 2,908 participants.

\subsubsection{Experiment}\label[appendix]{app::workflow}

We randomly allocated participants to experiment sessions. Participants could be invited to multiple sessions, with 7.2\% of participants attending more than one session. Participants were not eligible for re-invitation if they had completed the experiment or failed the quiz. In addition, participants were restricted to the same treatment condition if they had read the instructions. 

We messaged participants at least 24 hours in advance of the experiment start time. Experiment links were posted on Amazon Mechanical Turk at the exact time in the experiment invitation.

The workflow of the experiment is summarised in \cref{fig::experiment_workflow}. The experiment was coded in oTree~\parencite{Chen2016}. The mean submission time was 54.42 minutes (s.d. of 10.32 minutes). Participants were paid a fixed fee of \$2.00 plus any earnings from the market experiment and bonus tasks. The mean payment was \$12.81 (s.d. of \$4.70) inclusive of the fixed fee.

Participants completed a comprehension test after reading the experiment instructions. The comprehension test comprised three questions. The first question asked how many quality levels there were in the experiment. The second question asked participants to compute buyer earnings based on example values. The third question asked participants to compute seller earnings based on example values. Participants had three attempts to answer the quiz questions correctly. The quiz failure rate was 17.4\% with a mean attempt rate of 1.74 (s.d. of 0.84). 51.44\% of participant completed the quiz successfully within one attempt.

Participants proceeded to a wait page upon successful completion of the quiz. Participants had to click on a progress bar every 30 seconds to remain active on the page. Active participants waited on the page until a group of 7 players could be formed or until they exited the experiment. A group of 7 players could be formed once either: 10 participants were active on the wait page; or 1 participant had been waiting for more than 7 minutes and at least 7 participants were active on the wait page. Each participant could choose to leave the wait page after 10 minutes and receive the fixed fee of \$2.00. In addition, we set a maximum wait time of 20 minutes upon which participants automatically received the fixed fee. 16.2\% of eligible participants could not be assigned to a group for the experiment.

Participants assigned to a group proceeded to the market experiment described in the main text. Participants were paid the earnings from 7 randomly chosen market rounds plus a fixed endowment (200 points). The fixed endowment was designed to cover any losses arising during the market experiment. During the market experiment, participants had to wait on wait pages while others in their group made decisions. On each wait page, participants had to click on a progress bar every 30 seconds to avoid timing out. There were also timeouts on the decision pages. Participants could not continue with the experiment if they timed out on three consecutive pages. In addition, buyers received a bonus of 30 points if they did not timeout on any page in a market round. We introduced the attention bonus to reduce the gap in earnings between buyers and sellers. The attention bonus was awarded in 88.3\% of cases.

There were two bonus tasks after the market experiment. In the first bonus task, participants played the role of recipient in the Trust Game~\parencite{Berg1995}. We used the strategy method with ex-post matching. In the second bonus task, participants played the Bomb Risk Elicitation Task~\parencite{Crosetto2013}. Participants decided how many boxes to collect out of 100 boxes, one of which contained a bomb. Earnings increased linearly per box (\$0.02). Participants earned \$0.00 if they collected the box with the bomb.

We ran experiment sessions between 4 March 2022 and 21 April 2022. 791 participants took part in the market experiment.

\subsection{Data Collection}

We begin by defining our exclusion criterion. Next, we describe the final dataset. Finally, we present balance tests for the treatment conditions.

\subsubsection{Exclusion Criterion}

We restrict our sample to complete groups for which all participants completed the market experiment and the bonus tasks. Our design is sensitive to attrition due to the small number of buyers and sellers. 10.9\% of participants did not complete the market experiment and the bonus tasks, and 49.6\% of groups were complete. \Cref{tab::counts} presents the number of participants in our final sample, separated by treatment condition.

\begin{table}[htbp]
    \begin{center}
    \caption{Sample counts by treatment\label{tab::counts}}
    \begin{tabular}{c c c c c c}

        \toprule

        \multirow{2}{*}{Treatment} & \multirow{2}{*}{Rating attack} & \multirow{2}{*}{Delivery attack} & \multicolumn{3}{c}{\underline{Sample counts}} \\
            
        & & & Groups & Sellers & Buyers \\

        \midrule

        Baseline (B) & \ding{55} & \ding{55} & 14 & 42 & 56 \\

        Delivery (B) & \ding{55} & \ding{51} & 14 & 42 & 56 \\

        Rating (B) & \ding{51} & \ding{55} & 14 & 42 & 56 \\

        Combined (C) & \ding{51} & \ding{51} & 14 & 42 & 56 \\

        \bottomrule

    \end{tabular} \end{center}
    
\end{table}

\subsubsection{Summary Statistics}

We present summary statistics in \cref{tab::summary_stats}. 43.9\% of our sample is female. 21.2\% of our sample defines themselves as a race other than white. The average age is 38.7 years (s.d. of 10.75 years). The modal household income range is \$40,000 - \$59,999, with 33.7\% of participants in this category. 57.7\% of our sample is married and the average household size is 2.9 people (s.d. of 1.37). 75.3\% have completed a 4-year College degree or higher-level qualification. 35.5\% work in a ``Management, professional and related'' occupation. 69.4\% of participants live in a city with a population greater than 50,000. In the Trust Game, the average amount sent is \$0.54, with 18.9\% sending \$0.00 and 35.2\% sending \$1.00.

\subsubsection{Balance Tests}

We conduct non-parametric tests of treatment differences, using Fisher’s exact test for discrete variables, the Pearson chi-squared test for categorical variables and the Kruskal-Wallis test for continuous variables. We cannot reject treatment balance at a 5\% significance level for any of the demographic variables. P-values are reported in \Cref{tab::summary_stats}.

\begin{table}[htbp]
    \centering
    \caption{Descriptive statistics}\label{tab::summary_stats}
    \resizebox{0.65\textwidth}{!}{
    
    \begin{threeparttable}

    \begin{tabular}{ccccccc}
        \toprule
     Variable & All & Baseline ($B$) & Rating ($R$) & Delivery ($D$) & Combined ($C$) & p-value \\
     \hline
     Female & 0.439 & 0.388 & 0.398 & 0.480 & 0.490 & 0.343 \\
     & \small{(0.03)} & \small{(0.05)} & \small{(0.05)} & \small{(0.05)} & \small{(0.05)} &  \\ [0.5em]
     Age & 38.656 & 38.510 & 39.929 & 38.561 & 37.622 & 0.669 \\
     & \small{(0.543)} & \small{(1.12)} & \small{(1.12)} & \small{(1.06)} & \small{(1.04)} & \\ [0.5em]
    Non-white & 0.212 & 0.153 & 0.235 & 0.276 & 0.184 & 0.169 \\
    & \small{(0.02)} & \small{(0.04)} & \small{(0.04)} & \small{(0.05)} & \small{(0.04)} & \\ [0.5em]
     Household income & & & & & &  0.499 \\
     \small{Under \$20,000} & 0.066 & 0.031 & 0.092 & 0.082 & 0.061 & \\
     & \small{(0.01)} & \small{(0.02)} & \small{(0.03)} & \small{(0.03)} & \small{(0.02)} & \\
    \small{\$20,000 - \$39,000} & 0.186 & 0.245 & 0.194 & 0.153 & 0.153 & \\
     & \small{(0.02)} & \small{(0.04)} & \small{(0.04)} & \small{(0.04)} & \small{(0.04)} & \\
     \small{\$40,000 - \$59,999}  & 0.337 & 0.296 & 0.286 & 0.388 & 0.378 & \\
     & \small{(0.02)} & \small{(0.05)} & \small{(0.05)} & \small{(0.05)} & \small{(0.05 )} & \\
      \small{\$60,000 - \$ 79,999} & 0.161 & 0.194 & 0.133 & 0.122 & 0.194 & \\
     & \small{(0.02)} & \small{(0.04)} & \small{(0.03)} & \small{(0.03)} & \small{(0.04)} & \\
      \small{\$80,000 - \$99,999} & 0.105 & 0.133 & 0.082 & 0.102 & 0.102 & \\
     & \small{(0.02)} & \small{(0.03)} & \small{(0.03)} & \small{(0.03)} & \small{(0.03)} & \\
     \small{\$100,000 - \$119,999} & 0.074 & 0.051 & 0.122 & 0.071 & 0.051 & \\
     & \small{(0.01)} & \small{(0.02)} & \small{(0.03)} & \small{(0.03)} & \small{(0.02)} & \\
     \small{\$120,000 - \$149,999} & 0.031 & 0.010 & 0.05 & 0.03 & 0.03 & \\
     & & \small{(0.01)} & \small{(0.02)} & \small{(0.02)} & \small{(0.02)} & \\
     \small{\$150,000+} & 0.041 & 0.04 & 0.04 & 0.05 & 0.03 & \\
     & \small{(0.01)} & \small{(0.02)} & \small{(0.02)} & \small{(0.02)} & \small{(0.02)} & \\ [0.5em]
     Household size & 2.934 & 2.969 & 2.878 & 2.765 & 3.122 & 0.289 \\
     & \small{(0.07)} & \small{(0.12)} & \small{(0.14)} & \small{(0.14)} & \small{(0.15)} & \\ [0.5em]
     Married & 0.577 & 0.684 & 0.520 & 0.531 & 0.571 & 0.078 \\
     & \small{(0.02)} & \small{(0.05)} & \small{(0.05)} & \small{(0.05)} & \small{(0.05)} & \\ [0.5em]
     College degree & 0.753 & 0.837 & 0.714 & 0.745 & 0.714 & 0.137 \\
     & \small{(0.02)} & \small{(0.04)} & \small{(0.05)} & \small{(0.04)} & \small{(0.05)} &  \\ [0.5em]
     Professional & 0.355 & 0.316 & 0.327 & 0.388 & 0.388 & 0.602 \\
     & \small{(0.02)} & \small{(0.05)} & \small{(0.05)} & \small{(0.05)} & \small{(0.05)} &  \\ [0.5em]
     Urban & 0.694 & 0.704 & 0.694 & 0.673 & 0.704 & 0.976 \\
     & \small{(0.02)} & \small{(0.05)} & \small{(0.05)} & \small{(0.05)} & \small{(0.05)} &  \\ [0.5em]
     Census Bureau Regions & & & & & & 0.498 \\ 
     North East & 0.173 & 0.184 & 0.122 & 0.214 & 0.173 & \\
     & \small{(0.02)} & \small{(0.04)} & \small{(0.03)} & \small{(0.04)} & \small{(0.04)} \\ 
     Mid West & 0.173 & 0.173 & 0.184 & 0.173 & 0.163 & \\
     & \small{(0.02)} & \small{(0.04)} & \small{(0.04)} & \small{(0.04)} & \small{(0.04)} \\ 
     South & 0.372 & 0.347 & 0.357 & 0.408 & 0.378 & \\
     & \small{(0.02)} & \small{(0.05)} & \small{(0.05)} & \small{(0.05)} & \small{(0.05)} \\ 
     West & 0.281 & 0.296 & 0.337 & 0.204 & 0.286 & \\
     & \small{(0.02)} & \small{(0.05)} & \small{(0.05)} & \small{(0.04)} & \small{(0.05)} \\ [0.5em] 
     Trust & 0.542 & 0.556 & 0.557 & 0.519 & 0.535 & 0.895 \\
     & \small{(0.02)} & \small{(0.04)} & \small{(0.04)} & \small{(0.04)} & \small{(0.04)} & \\
     \hline
     No. observations & 392 & 98 & 98 & 98 & 98 \\
     \bottomrule 
    \end{tabular}

\begin{tablenotes}
    \small
    \item \footnotesize Notes: The table contains sample means with standard errors in parentheses. P-values correspond to non-parametric tests of treatment differences. We use Fisher's exact test for discrete variables, the Pearson chi-squared test for categorical variables and the Kruskal-Wallis test for continuous variables. 
\end{tablenotes}

\end{threeparttable}
}
\end{table} 

\clearpage

\section{Experiment instructions}\label[appendix]{app::instructions}

In \cref{app::general_instructions}, we present the instructions shown to all participants at the beginning of the experiment. In \cref{app::role_specific_instructions}, we present the role-specific instructions shown to participants at the beginning of the market experiment. 

\subsection{General instructions}\label[appendix]{app::general_instructions}
\noindent \underline{Page One} \\

\noindent Welcome and thank you for participating in this experiment run by researchers at the University of Cambridge (UK) and the University of Oxford (UK). \\

\noindent The research study has received ethical approval from the Research Ethics Committee of the Department of Sociology (DREC) at the University of Oxford. The DREC reference number is \textbf{SOC\_R2\_001\_C1A\_21\_56}. \\

\noindent The aim of this experiment is to study how individuals make decisions in certain contexts. In the experiment, you will interact with other Turkers in real time. You will make decisions that will affect the amount of money you earn and the amount of money other Turkers earn. \\

\noindent It is important for you to know that your decisions will remain completely \textbf{confidential}. We will never reveal your Turker ID or any other information that might allow other participants to identify you. \\

\noindent \underline{Page Two} \\

\noindent The expected duration of the experiment is around \textbf{40 minutes}. \\

\noindent There are four parts of the experiment. 

\begin{itemize}
    \item \textbf{Instructions for Task One}
    \item \textbf{Quiz on instructions}
    \item \textbf{Task One: decision-making task}
    \item \textbf{Bonus tasks}
\end{itemize}

\noindent You must pass the quiz to participate in Task One and the Bonus tasks. You will have \textbf{three} attempts to pass the quiz. \\

\noindent Turkers who pass the quiz will be allocated to a group for Task One after the quiz. \\

\noindent \underline{Page Three} \\

\noindent \textbf{Earnings} \\

\noindent You will earn a fixed \textbf{\$2.00} fee for sure if you pass the quiz and complete the experiment. \\

\noindent In addition, you may earn a bonus payment. We expect the average total earnings to be within the \textbf{\$9-\$12} range. \\

\noindent After you complete the experiment, we will send you the payment within 48 hours. \\

\noindent \textbf{If you fail the quiz or exit the experiment before completion, you will not receive any earnings.} \\

\noindent Please click on the next button below to continue to the instructions for Task One. \\

\noindent \underline{Page Four} \\

\noindent Task One is a market experiment. Turkers participate in a market for \textbf{20 trading periods}. \\

\noindent \textbf{Market} \\

\noindent Each market contains \textbf{seven Turkers}, who are divided into buyers and sellers. There are \textbf{three sellers} and \textbf{four buyers} in each market. Each Turker's role stays the \textbf{same} throughout the market. \\

\noindent Sellers are labelled by upper-case letters, e.g., \emph{Seller M}. Buyers are labelled by lower-case letters, e.g., \emph{Buyer f}. \\

\noindent You will be allocated a role (either buyer or seller) at the start of Task One. \\

\noindent There are \textbf{two} qualities of good: Regular and Super. \\

\noindent \underline{Page Five} \\

\noindent \textbf{Sellers} \\

\noindent In each trading period, each seller decides:
\begin{itemize}
    \item How many goods to produce
    \item How many goods to advertise
    \item A price for the goods.
\end{itemize}

\noindent \textbf{Production} \\

\noindent Each seller picks \textbf{one} of the following options for production:
\begin{itemize}
    \item 1 Regular good
    \item 1 Super good
    \item 2 Regular goods
    \item 2 Super goods
    \item Nothing.
\end{itemize}
\noindent It costs \textbf{more} for a seller to produce a Super good than a Regular good. It costs \textbf{more} for a seller to produce the second good than the first good. Sellers will learn the exact costs of production at the start of Task One. \\

\noindent A seller must pay the costs of production, even if their goods are not sold. \\

\noindent \textbf{Advertising} \\

\noindent A seller decides how many goods to advertise for sale. Buyers can only buy advertised goods and do not know how many goods each seller produces. \\

\noindent Notice that advertisements do not have to be truthful. A seller can advertise more goods than produced and goods of a \textbf{different} quality than produced. 

\begin{center}
    \noindent << The following text is only shown in the Delivery and Combined treatments >>
\end{center}

\noindent There is a \textbf{20\%} probability that each good is not delivered. For example, if Seller M produces 2 Regular Goods and sells one of these goods to Buyer f, there is a 20\% probability that Buyer f receives \textbf{nothing}. \\

\noindent Note that a good may not be delivered either because the advert is not truthful or because of the 20\% probability that the good does not get delivered. Buyers will not be told the reason a good is not delivered. Sellers will not be told whether a good is not delivered.

\begin{center}
    \noindent  << - - - >>
\end{center}

\noindent \textbf{Earnings in a trading period} \\

\noindent A seller's earnings equal their income from sales less the costs of production. 

\begin{center}
    \noindent Seller Earnings=(Number of Goods Sold $\times$ Price)-Costs of Production
\end{center}

\noindent \textbf{Example} \\

\noindent Seller M produces 2 Regular Goods. Imagine, as an example, that it costs A points to produce the first good and B points to produce the second good. This means that the sum of the costs of production is A points + B points = C points. \\

\noindent Seller M advertises 2 Super Goods and sets a price of D points for each good. \\

\noindent Seller M sells one good. Seller M receives (1 $\times$ D points) = D points from sales. Seller M's earnings equal the income from sales (D points) less the costs of production (C points).

\begin{center}
    \noindent Seller M' s Earnings=(1 $\times$ D points)-C points=E points
\end{center}

\noindent \underline{Page Six} \\

\noindent \textbf{Buyers} \\

\noindent In each trading period, each buyer decides whether to buy \textbf{one} of the available goods. A buyer can choose to \textbf{not} buy a good. \\

\noindent Buyers make their purchase decisions in order. The first buyer decides whether to buy one of the available goods, then the second buyer and so on. If a buyer purchases a good, that good \textbf{cannot} be purchased by subsequent buyers. \\

\noindent At the time of purchase, the buyer does \textbf{not} know the quality of the good. The buyer only knows the advertised quality. \\

\noindent The advertised quality does not have to equal the actual quality. Recall that advertisements do not have to be truthful. A seller can advertise more goods than produced and goods of a different quality than produced. \\

\noindent The quality of the good is revealed \textbf{after} purchase. \\

\noindent \textbf{Valuations} \\

\noindent Buyers value Super goods \textbf{more} than Regular goods. Buyers will learn the exact valuations at the start of Task One. \\

\noindent \textbf{Ratings}  \\

\noindent If a buyer purchases a good, they must rate the seller on a five-point numerical scale after the quality of the good is revealed. On the scale, 5 corresponds to Very Good and 1 corresponds to Very Poor. 

\begin{center}
    \noindent << The following text is only shown in the Rating and Combined treatments >>
\end{center}

\noindent There is a \textbf{20\%} probability that each rating is replaced with a random \textbf{different} rating. For example, if Buyer f submits a rating of 5 for Seller M, there is a 20\% probability that Buyer f's rating is replaced with 1, 2, 3 or 4. Turkers will \textbf{not} be told if a rating has been replaced.

\begin{center}
    << - - - >>
\end{center}

\noindent Each seller's rating is updated at the end of each trading period. \\

\noindent We use `?' to signal that a rating does not exist. In the first trading period, each seller's average rating is `?' and each seller's last three ratings are [`?', `?', `?']. \\

\noindent \textbf{Information}  \\

\noindent At the time of purchase, a buyer can see for each available good:
\begin{itemize}
    \item The advertised quality (i.e., Regular or Super)
    \item The price of the good
    \item The seller's average rating
    \item The seller's last three ratings
    \item The seller's unique identification letter.
\end{itemize}

\noindent \textbf{Earnings in a trading period} \\

\noindent If a buyer purchases a good, the buyer's earnings equal the \textbf{value of the good less the price of the good}. \\
\begin{center}
     Buyer Earnings=Value of Good-Price Paid
\end{center}

\noindent If a buyer does not purchase a good, the buyer earns \textbf{0 points}. \\

\noindent \textbf{Example} \\

\noindent Following on from the example on the previous page, Buyer f purchases a Super good from Seller M for D points. \\

\noindent At the time of purchase, Seller M's last three ratings are [`?', F, G] and Seller M's average rating is H. \\

\noindent Buyer f receives a Regular good from Seller M. The advertised quality (Super) does not equal the actual quality (Regular). In this example, the advert is not truthful. \\

\noindent Buyer f gives Seller M a rating of I after the quality of the good is revealed. In the next trading period, Seller M's last three ratings will be [F, G, I] and Seller M's average rating will be J. \\

\noindent Imagine, as an example, that Buyer f values a regular good at K points. Buyer f's earnings equal the value of the good (K points) less the price of the good (D points). 
 
\begin{center}
    Buyer f's Earnings=K points-D points=L points
\end{center}

\noindent \underline{Page Seven} \\

\noindent \textbf{Earnings} \\

\noindent In Task One, you will be paid a fixed fee of \textbf{200 points} plus the earnings from \textbf{seven} randomly chosen trading periods. \\

\noindent At the end of the experiment, your points will be converted to dollars at a fixed conversion rate. \textbf{100 points is equivalent to \$2.00}. \\

\noindent Recall, you will also earn a show-up fee of \textbf{\$2.00} if you pass the quiz and complete all parts of the experiment, including the Bonus Tasks. You may also earn bonus payments in the Bonus Tasks. We expect the average total earnings to be within the \textbf{\$9-\$12 range}. \\

\noindent You must remain \textbf{active} during Task One. If you timeout on \textbf{three consecutive pages}, you will not receive any earnings. You will be informed if you timeout during Task One. \\

\noindent If you exit the experiment at any point before completion, you will \textbf{not} receive any earnings. You will be informed if you timeout during Task One. \\

\noindent Note that you will be able to continue the experiment in case one or more of the Turkers in your group drop out for technical reasons. You will be informed if a Turker drops out of the experiment. \\

\noindent Please click on the next button to proceed to the quiz. You must pass the quiz to participate in Task One and the Bonus Tasks. You will have \textbf{three} attempts to pass the quiz. \\

\noindent IMPORTANT. You should not refresh your browser or press the back button at any point during the quiz or Task One. \\

\subsection{Role-specific instructions}\label[appendix]{app::role_specific_instructions}
\subsubsection{Buyers}

\noindent \textbf{Valuations} \\

\noindent The value of a Regular good is \textbf{30 points}. The value of a Super good is \textbf{150 points}. You will be able to access the good valuations and the Task One instructions during the experiment. \\

\noindent \textbf{Timeouts}

\noindent There are timeouts on each page. \\

\noindent In \textbf{each} trading period, you will earn a bonus of \textbf{30 points} if you do not timeout on any page during that trading period. The bonus does not depend on whether you purchase a good. You will be informed whether you have earned a bonus at the \textbf{end} of each trading period. \\

\noindent Recall that you will be paid the earnings from \textbf{seven randomly chosen} trading periods. You will only earn the bonus payment if that trading period is selected for payment. \\

\noindent If you timeout on \textbf{three consecutive pages}, you will \textbf{not} receive any earnings. \\

\noindent Please click on the next button to proceed to the experiment. \\

\subsubsection{Sellers}
\noindent \textbf{Costs of production} \\

\begin{center}
    \begin{tabular}{c | c | c}
        \hline 
    & Regular & Super \\
    \hline 
First good & 10 points & 50 points \\
Second good & 20 points & 100 points \\
\hline 
\end{tabular}
\end{center}

\noindent You \textbf{must} pay the costs of production, even if your goods are not sold. \\

\noindent \textbf{Sum of costs of production} \\

\noindent We have calculated the sum of the costs of production for each production option in the table below. \\

\begin{center}
    \begin{tabular}{c|c}
        \hline 
    Production decision & Sum of costs of production \\
    \hline 
    1 Regular good & 10 points \\
    1 Super good & 50 points \\
    2 Regular goods & 10 points + 20 points = 30 points \\
    2 Super goods & 50 points + 100 points = 150 points \\
    Nothing & 0 points \\
    \hline 
\end{tabular}
\end{center}

\noindent You will be able to access the costs of production and the Task One instructions during the experiment. \\

\noindent There are timeouts on each page. If you timeout on \textbf{three consecutive pages}, you will \textbf{not} receive any earnings. \\

\noindent Please click on the next button to proceed to the experiment. \\

\end{document}